# A model of mentorship for students from historically underrepresented groups in STEM

M. K. Rodriguez Wimberly[1†], Alexander L. Rudolph[2], Carol Hood[3], Rachel E. Scherr[4], and Christine Pfund[5]

[1] University of California, Riverside, CA 92507
[2] California State Polytechnic University, Pomona, CA 91768
[3] California State University, San Bernardino, CA 92407
[4] University of Washington, Bothell, WA 98011
[5] Center for the Improvement of Mentored Experiences in Research (CIMER), Wisconsin Center for Education Research, University of Wisconsin-Madison, Madison, WI 53705
[†] National Science Foundation Mathematics and Physical Sciences Postdoctoral Ascend Fellow

## Keywords

Stem; undergraduates; underrepresented minorities; mentoring; student support programs; programmatic mentoring

## Abstract

Mentorship is critical to student academic success and persistence, especially for students from historically underrepresented (HU) groups. In a program designed to support the academic success of HU undergraduates in STEM who wish to pursue a PhD in those fields, students experience comprehensive support including financial aid, highly-engaged mentoring, dual faculty mentorship, professional development workshops, and summer research experiences. Scholars in this program, the Cal-Bridge program, consistently report that faculty mentorship is the most impactful feature. While mentorship was rated highly, preliminary evaluation indicated an early deficit in a sense of community among scholars. In response, faculty professional development and support for peer networking were implemented to expand and enhance the relationships that support scholar success. Here we present a promising multifaceted model of mentorship that can support the academic success of HU undergraduates.

# Introduction: The Importance of Mentorship to Academic Success

In the 2019 report of the National Academy of Science, Engineering, and Medicine (NASEM) entitled, "The Science of Effective Mentorship," mentorship is defined as "a professional, working alliance in which individuals work together over time to support the personal and professional growth, development, and success of the relational partners through the provision of career and psychosocial support" (National Academies of Sciences, 2019).This definition highlights the critical role both mentor and mentee play in the effectiveness of the relationship. Research indicates that the frequency and quality of mentee-mentor interactions positively correlates with students' persistence in STEM degree programs. Mentorship has also been positively associated with students' identity, research self-efficacy, and their sense of belonging. For students from traditionally underrepresented racial and ethnic groups, mentorship has been positively correlated with enhanced recruitment into graduate school and research-related career pathways (see National Academies of Sciences, 2017, 2019; and Pfund, 2018 for a summary of the literature).

Importantly, the ability of mentors to meet the psychosocial needs of their mentees is associated with increases in how mentees perceive the quality of the mentoring relationship and satisfaction with that relationship (H. R. Tenenbaum et al., 2001; Waldeck et al., 1997). Additionally, when mentors have skills in cultural awareness they can better help students from underrepresented groups navigate the often challenging experience they face in majority environments and reinforces their research self-efficacy (Byars-Winston et al., 2015). In addition, quality mentorship increases HU students' sense of belonging, disciplinary identity development, and overall confidence to be a scientist (Estrada et al., 2018).

Mentoring relationships can occur in many forms including the traditional dyad, single mentors working with multiple mentees, and groups of mentors working with a single mentee as a mentorship network. "Mentorship networks—the constellations of mentoring relationships and resources that a mentee taps for support—have gained increasing recognition both within and outside of science, technology, engineering, mathematics and medicine (STEMM)" (National Academies of Sciences, 2019). Peer mentoring relationships, especially within a well-defined community, can also provide very powerful support. Thus, mentorship roles can be fulfilled by many individuals who engage with a given mentee including research advisors, instructors, program directors, program meeting leaders, peers, committee members, etc. (Rath et al., 2018).

In a program designed to support the academic success of HU undergraduates in STEM who wish to pursue a PhD in those fields, called the Cal-Bridge Program, students experience comprehensive support including highly-engaged mentoring, dual faculty mentorship, financial aid, professional development workshops, and summer research experiences. To investigate how various elements of the Cal-Bridge program influence (or not) retention and advancement, we conducted annual evaluations of the program, consisting primarily of interviews and surveys of scholars. Scholars in this program consistently report that faculty mentorship had the greatest impact of all the program elements. Scholars also described an early deficit in a sense of community among scholars, in response to which Cal-Bridge developed a peer mentoring program.

The remainder of the article is broken into five parts. First, we describe the Cal-Bridge program, with an emphasis on the "multifaceted highly-engaged mentorship" model. We then describe the initial evaluation of the program. The third section discusses the findings of the initial evaluation with focus on the positive impact from multifaceted and highly-engaged faculty mentoring. The following section highlights applications of the evaluation's findings, namely the creation of a peer mentorship program and major revision of the program's mentorship handbook. In sum, we describe a promising multifaceted model of mentorship that can support the academic success of HU undergraduates.

# Context for Study: The Cal-Bridge Program

Our study takes place in the context of the Cal-Bridge Program (www.calbridge.org) , an NSF-funded program that helps current STEM students in California State Universities (CSUs) and California Community Colleges (CCCs) attain their undergraduate Bachelor's degree and successfully apply to graduate school (Rudolph, 2019; Rudolph et al., 2019). Since its inception in 2014, the main Cal-Bridge program has had great success, with 72% of the scholars who applied ($N$=127) accepted to a PhD program; another remaining 18% are enrolled in MS programs with plans to attend a PhD program. At the time of writing 2 scholars from the first cohort of 5 have received their PhD.

The Cal-Bridge program is a partnership between all three tiers of the California higher education system: 9 University of California (UC), 23 CSU, and 116 community college campuses in California, with more than 300 faculty from the three systems participating. A second subprogram, called Cal-Bridge Summer (previously known as CAMPARE; www.cpp.edu/calbridge/summer-research), is designed to help Cal-Bridge scholars as well as other CSU and community college students gain access to a wide variety of summer research opportunities at more than 20 leading research institutions across the country. The current Cal-Bridge program originally focused on physics and astronomy; more recently the program has added computer science/computer engineering and mathematics and statistics to the disciplines from which scholars are drawn; eventually, the program will expand to include other STEM fields.

The philosophy of the Cal-Bridge program is based on the concept of high expectations coupled with the support required to meet those expectations. The Cal-Bridge program uses research-validated selection methods to identify HU students who display strong socio-emotional competencies, along with academic potential. Once a student is selected into the Cal-Bridge program, they receive that support through four "pillars of support" (Figure 1): multi-level mentorship, financial aid, professional development workshops, and summer research experience opportunities. As a result of this support many scholars successfully matriculate to a PhD program, including to the UC campuses in the Cal-Bridge network, which will be discussed in detail in Section 4 on the evaluation's findings.

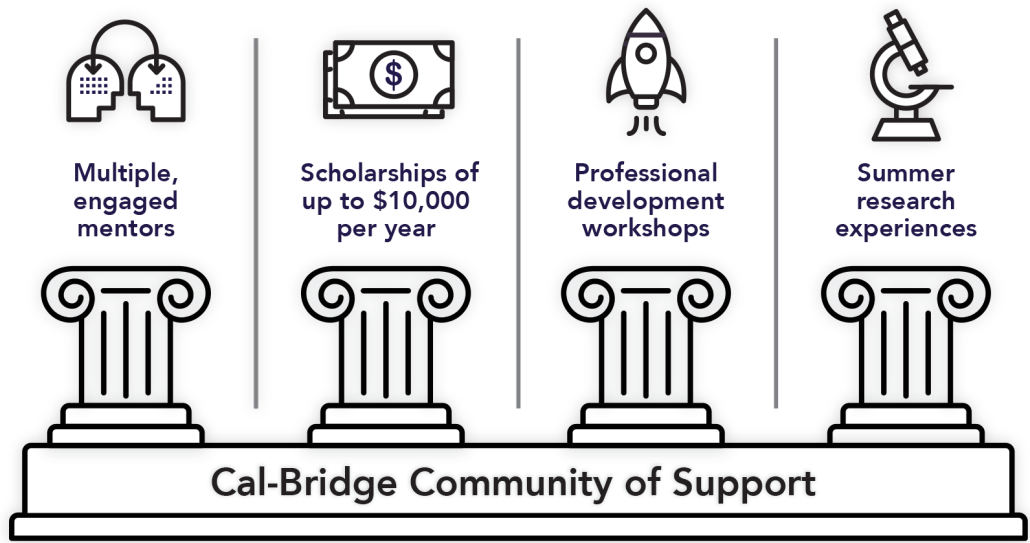

*Figure 1. The Cal-Bridge four pillars of support undergirded by a community of scholars, alumni, faculty, and staff*

## First Pillar: Multiple, Highly-Engaged Mentors

The Cal-Bridge faculty mentors are the most frequent contact each scholar has with the program. Scholars meet with their mentors every other week during the academic year to discuss their academic progress, their well-being, and to provide advice on the scholar's PhD program and summer research experience applications preparation and selection of any resulting offers. Frequent meetings ensure that issues scholars may have do not go unaddressed for very long.

Each scholar is assigned two faculty mentors, once from their home CSU campus and a second from a UC campus. Mentors are assigned  such that at least one mentor, and often both, have a research background in the scholar's proposed research area (physics, astronomy, computer science/engineering or math). In most cases, at least one mentor is of the same gender as the scholar; less frequently, but whenever possible, at least one mentor is assigned from the same racial, ethnic, or cultural background as the scholar.

Mentors are given access to training and guidance to provide culturally aware mentorship for scholars to feel comfortable talking about not only academic issues but also other concerns that might be impacting their academic success, including mental health issues, home or family issues, or other personal matters. Cal-Bridge mentors are supported in this highly-engaged mentorship practice in 3 ways:

1. A Mentorship Handbook[1], written by experienced mentors and former scholars from within the Cal-Bridge program to address both mentor and scholar responsibilities and best practices.
2. A series of professional development workshops on best practices in mentorship, including mentoring students from backgrounds different from that of the mentor (Butz et al., 2018; Handelsman et al., 2005; National Academies of Sciences, 2019; Pfund et al., 2006, 2014, 2015; Weber-Main et al., 2019)[2].
3. Additional mentorship workshops led by national experts in mentoring, including The Science of Effective Mentoring (National Academies of Sciences, 2019) and Mentoring for Wellbeing[3].

# Second pillar: Financial aid

For students with low socioeconomic status, academic success may be determined by the amount of time they are able to spend on their studies, which is severely limited by their need to work at a non-academic job while in school. Before becoming Cal-Bridge scholars, almost half of the students in the program were working more than 20 hours a week at non-academic jobs. Financial aid provided by the Cal-Bridge program allows students to reduce or eliminate their outside work hours so that they can focus on school. Once in the program, scholars are limited to working a maximum of 10 hours per week when classes are in session.

Financial aid is provided to scholars based on their federally determined financial need, up to $10,000 a year.[4] The average award for scholars who receive aid is just under $8000, and over 80% receive more than $5000. About a quarter (23%) of scholars do not receive aid, most because they do not have financial need, but others because they are not eligible due to their citizenship status (DACA/California AB540 or international). The program raises private donations to support these other scholars.

---

[1] This handbook was originally written by a few experienced Cal-Bridge faculty mentors and members of program leadership. Due to the evaluation results discussed in this work the handbook was completely revised by former scholars and program leaders. The newly implemented version is discussed in Section 5. This current version can be found here: https://tinyurl.com/CBMentorshipHandbook.

[2] These workshops and more resources can be found on the American Physical Society's Mentor Resource Library (APS MRL, https://aps.org/programs/minorities/nmc/webinars.cfm).

[3] Specifically a workshop related to "Fostering Independence with Growth Mindset and Mental Health" linked in the APS MRL.

[4] This $10,000 limit is imposed by the terms of the NSF grant program currently funding the program. In future, with sufficient funding, the program would provide 100% of each scholar's unmet financial need.

## Third pillar: Professional development workshops

Approximately once a month during the academic year, scholars attend professional development workshops on a wide range of topics including socio-emotional competencies (e.g., imposter syndrome, harassment, and growth mindset), Python programming, applying to summer research experiences, preparing for graduate school, writing graduate application essays, and applying for fellowships. These workshops are held either virtually or on a UC campus: in-person workshops include opportunities to tour campus facilities and meet faculty and graduate students. These workshops also connect scholars and faculty from multiple campuses.

## Fourth pillar: Summer research experiences

Summer research experiences are one of the most important mechanisms for scholars to confirm their commitment to applying and attending graduate school, and for graduate admissions committees to select scholars for admission (National Academies of Sciences, 2017; Russell et al., 2007; Seymour et al., 2004). The Cal-Bridge program runs a summer research experience program, called Cal-Bridge Summer, that matches California community college and CSU students to research scientists across the country, and about two-thirds of Cal-Bridge scholars are placed through this matching program and engage in 8-10 weeks of full-time research. Summer research advisors provide an additional layer of mentorship to Cal-Bridge scholars, including advising for scholars applying to graduate school along with often providing letters of reference for graduate school applications.

# Methods

The Cal-Bridge program employs an external, independent evaluator (author Scherr) who is responsible for conducting annual evaluations for the purposes of formative assessment of the program. Assessment is done through a combination of surveys and interviews, querying both scholars and mentors. Scholars were informed that responses to surveys and interviews are confidential: no member of the Cal-Bridge team has access to information on who gave which responses, except the independent program evaluator. These assessments were done under Cal Poly Pomona IRB protocol #18-186. This section outlines the details of the methodology of the scholar surveys and interviews from two recent years of the program, 2018-19 and 2019-20.

## Participants

Participants are undergraduate students in the CSU; many are transfer students from the CCC system. The CSU is the largest, most diverse public university system in the country, with over 430,000 undergraduates at 23 campuses, 47% of whom are from HU groups (NCES 2018). Twenty-one (21) of the 23 CSU campuses are Hispanic Serving Institutions or HSIs (>25% Hispanic enrollment), including 9 Minority Serving Institutions or MSIs (>50% HU student enrollment). Thus, the CSU system represents a large and diverse pool from which to draw promising scholars with the potential to matriculate to a STEM PhD program. The 116 CCCs, the majority of which are MSIs/HSIs, are the entry point into higher education for a large number of California students, particularly those from many HU groups. As many as half of the students at CSU campuses have transferred from community colleges.

Cal-Bridge scholars are very diverse: of 182 scholars in 8 cohorts from 2014-2021, 107 are members of a HU group, underrepresented racial and ethnic groups, including 87 Latine[5], 21 Black, and 7 Indigenous scholars.[6] 72 are women; notably, 30 of these women (17% of all scholars), are from historically underrepresented racial and ethnic groups in STEM; 112 scholars are first-generation college students, 34 identify as LGBTQ+, and 23 indicate they have a disability. All but one (1) of the 182 Cal-Bridge scholars (>99%) fall into one or more of these categories of HU students. Table 1 lists the demographics for each cohort in the program from 2014-2021.

**Table 1. Demographics of Cal-Bridge Scholars**

| Year of entry | 2014 | 2015 | 2016 | 2017 | 2018 | 2019 | 2020 | 2020 | 2021 | 2021 | | |
|---|---|---|---|---|---|---|---|---|---|---|---|---|
| Cohort | 1 | 2 | 3 | 4 | 5 | 6 | 7 P&A* | 7 CS* | 8 P&A* | 8 CS* | Total | Overall % |
| Total | 5 | 7 | 10 | 12 | 25 | 40 | 30 | 8 | 35 | 10 | 182 | |
| Man | 1 | 7 | 6 | 7 | 13 | 19 | 17 | 6 | 23 | 5 | 104 | 57.1% |
| Woman | 4 | 0 | 4 | 5 | 12 | 19 | 13 | 1 | 9 | 5 | 72 | **39.6%** |
| Nonbinary | 0 | 0 | 0 | 0 | 0 | 2 | 0 | 0 | 3 | 0 | 5 | **2.7%** |
| Decline | 0 | 0 | 0 | 0 | 0 | 0 | 0 | 1 | 0 | 0 | 1 | 0.5% |
| | | | | | | | | | | | | |
| Hispanic | 3 | 5 | 6 | 8 | 12 | 20 | 15 | 2 | 13 | 3 | 87 | 47.8% |
| Black | 0 | 0 | 3 | 2 | 5 | 2 | 2 | 1 | 4 | 2 | 21 | 11.5% |
| Amind | 1 | 1 | 0 | 2 | 1 | 1 | 0 | 0 | 1 | 0 | 7 | 3.8% |
| PacIs | 0 | 0 | 0 | 2 | 0 | 1 | 1 | 2 | 0 | 2 | 8 | 4.4% |
| Asian | 0 | 2 | 0 | 2 | 4 | 5 | 3 | 4 | 2 | 4 | 26 | 14.3% |
| White | 4 | 1 | 1 | 3 | 8 | 17 | 10 | 1 | 11 | 1 | 57 | 31.3% |
| | | | | | | | | | | | | |
| URM** | 4 | 5 | 9 | 9 | 16 | 21 | 17 | 3 | 18 | 5 | 107 | **58.8%** |

[5] Latine is a gender neutral term replacing Latino/Latina. Many people use Latinx for this purpose, but many native Spanish speakers object that Latinx is difficult to pronounce in Spanish and prefer Latine.
[6] The numbers of scholars from individual historically underrepresented racial and ethnic groups (URM groups) do not add up as some scholars hold multiple URM identities.

| | | | | | | | | | | | | |
|---|---|---|---|---|---|---|---|---|---|---|---|---|
| URM women | 3 | 0 | 3 | 4 | 6 | 7 | 7 | 0 | 3 | 2 | 30 | **16.5%** |
| First Gen | 4 | 4 | 10 | 11 | 14 | 23 | 19 | 5 | 17 | 5 | 112 | **61.5%** |
| LGBTQ+ | 0 | 2 | 1 | 0 | 5 | 11 | 4 | 0 | 8 | 3 | 34 | **18.7%** |
| Disability | N/A | N/A | N/A | N/A | N/A | 8 | 5 | 2 | 6 | 2 | 23 | **18.7%** |
| | | | | | | | | | | | | |
| URM or woman or first gen | | | | | | | | | | | 164 | 90.1% |
| HU*** | | | | | | | | | | | 181 | 99.5% |

*P&A refers Physics and Astronomy; CS includes Computer Science and Computer Engineering

**Underrepresented Minority (URM) groups specifically refer to the racial and ethnic groups that are underrepresented in STEM fields, namely Black, Native American, and Hispanic.

**Historically Underrepresented (HU) groups including students in underrepresented racial and ethnic groups, women, first generation, LGBTQ+, disabled; N/A refers to the fact that these data were not collected in that year.

# Surveys

Scholars were requested to complete surveys three times: once in June 2018, once in March-April 2019, and once in April 2020. For each survey, all scholars in the program at that time were requested to complete the survey; i.e., each year the survey was administered to new scholars as well as all returning scholars and alumni. Surveys were intended to gather formative feedback about the Cal-Bridge program based on scholars' experiences. Surveys were designed, administered, and analyzed by the program evaluator. Survey data included in this paper is based on responses to the following two questions:

1. What are the best parts of the Cal-Bridge program? (open-ended text response)

2. Please indicate the quality of the following Cal-Bridge elements (multiple choice responses: poor, fair, good, very good, excellent):
   - Providing financial aid
   - Faculty mentoring
   - Connecting to other Cal-Bridge scholars
   - Providing workshops and programs
   - Preparing a path for scholars to attend graduate school

In 2018, 31/34 (91%) scholars responded; in 2019, 37/59 (63%) scholars responded; and in 2020, 54/99 (54%) scholars responded (low response rate likely influenced by the COVID-19 pandemic); the combined response rate was 63.5%. Responses to the open-ended question were coded by the evaluator for emergent themes.

Surveys administered in this study had a high response rate, but bias may still be present: in particular, those with extreme experiences (highly positive or highly negative) may have been more likely to respond to the survey. Surveys did not define key concepts such as "community," meaning that responses may represent a variety of interpretations of such terms.

# Interviews

Interviews were conducted with Cal-Bridge scholars to gather formative feedback about elements of the Cal-Bridge program. Fourteen (14) scholars were interviewed in May-June 2018, 9 in Jan-Feb 2019, and 7 in March 2020, all from the physics and astronomy subprogram. Interviewees were selected to represent the overall characteristics of the Cal-Bridge scholar population including gender, race/ethnicity, financial aid status, and whether they are in the North or South part of the Cal-Bridge program. In 2018, interviewees were selected to represent the population of students in different cohorts of the program; in 2019 and 2020, interviewees were selected from the most recent cohort. Interviews took 10-30 minutes each and were conducted by videoconference, recorded, and transcribed for analysis. Figure 2 shows an abbreviated version of the interview protocol used in 2020. Interviews in 2018 and 2019 used only questions 1-3. The interviews were semi-structured; the protocol provided guidance for the interview, but the questions were not always asked precisely in these words or in this order. For example, if interviewees identified their race or ethnicity in response to questions 1-3, the interviewer did not ask for it again in question 5.

1. How is school going for you?
2. How is your life outside of school?
3. You may have interactions with people who have mentored or advised you. Could you tell me who those people are for you? What do you talk about? What do you get from this mentoring, or, what does this mentor do for you? Do you have a trusting relationship with your mentor?
4. Do you consider any other Cal-Bridge scholars to be "peer mentors" for you? Have you participated in the peer-mentoring program?
5. One goal of the Cal-Bridge program, as you may know, is to increase the number of physics PhDs from underrepresented groups. Do you identify as a member of an underrepresented group in physics? How would you describe yourself? [Answer = descriptor]
   a. Do you think that being [descriptor] matters for your experience in physics / astronomy? If so, how?
   b. (If yes:) Do you think that being [descriptor] matters for your mentoring relationships? If so, how?
   c. Are there any other [descriptor] students in your department (that you know of)?
   d. Are you connected with a group of [descriptor] in physics? How/where?
6. What would you want the Cal-Bridge program to know about how things are going for you?

*Figure 2. Abbreviated scholar interview protocol.*

Interviews were transcribed for analysis. Each transcript was systematically examined from beginning to end to discern (1) whether students had positive mentorship experiences, (2) whether the scholar reported an association between their mentorship experience and their social identity (e.g., racial identity, first-generation status), and (3) what mentorship behaviors scholars reported as being particularly helpful. Interviews were coded with these three codes, each of which could potentially appear multiple times in each interview. For each code that was assigned, a student quotation was identified that illustrated the code. Data associated with these three codes appears in the Findings: Faculty Mentorship section. Not every coded quotation is reported in this manuscript: the interviewer selected quotations that illustrate the themes most eloquently and completely.

Social and political context influence any interview; whether participants and interviewers have similar or different social identities, their positionality influences both what interviewees say and what interviewers (and researchers) perceive (Sultana, 2007). In this study, a white woman who is a senior academic professional interviewed junior people of color about their experiences, including racialized experiences, in a potentially stressful professional setting (physics undergraduate education). To help create an environment with the best opportunity for shared understanding, the interviewer took a number of steps outlined along with the interview protocol in Appendix A.

# Findings: Faculty mentorship

Survey findings indicate that mentorship is scholars' most favored element of the Cal-Bridge program: when asked, "What are the best elements of the Cal-Bridge program?" as an open-ended question, seventy-two percent (72%) of scholars included mentorship in their response ($N$=90), and mentorship accounted for 44% of all responses ($N$=154; see Figure 2).

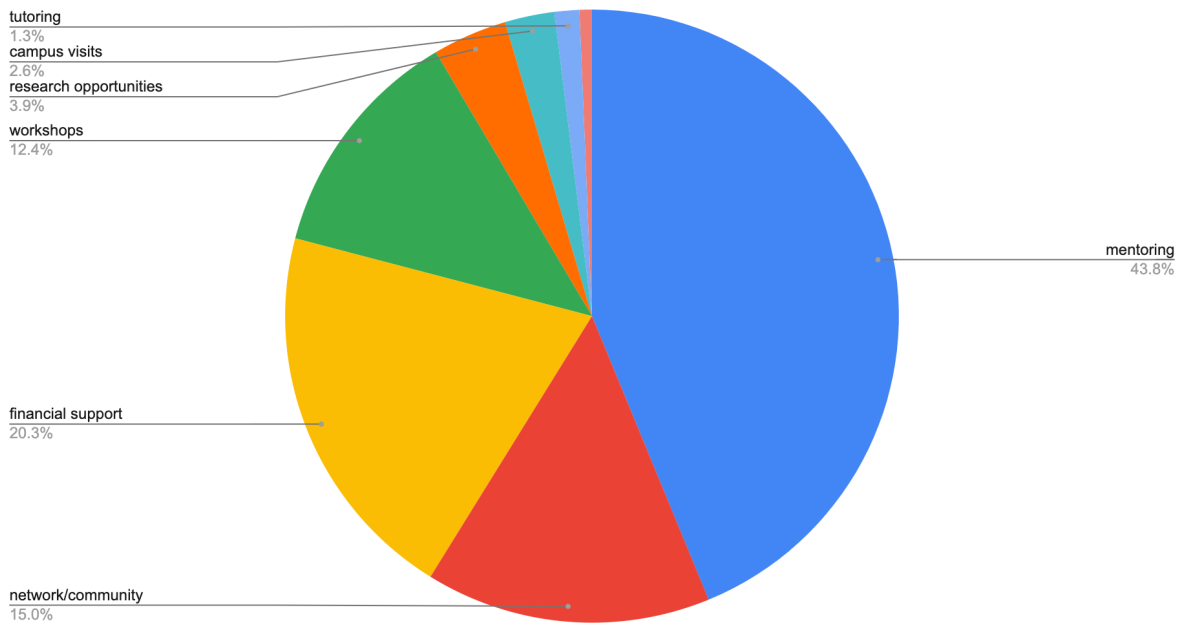

*Figure 3. Pie chart of responses to the question, "What are the best parts of the Cal-Bridge program?" The data represent the combined responses from 2019 and 2020 from a total of 90 respondents.*

Analyses of interviews and responses to open-ended survey questions confirm and enrich these findings:

> "Mentorship is really the main thing that helped me. There was financial support and workshop technical support, all of that. But I think really for me, I didn't have much mentorship before the end of my undergraduate career, so I think that made the major difference for me." (alumni graduate student, 2018)

In general, **scholars reported highly positive and supportive interactions with their Cal-Bridge mentors**:

> "Oh, they're super awesome. They're really encouraging. They're always looking for ways to help. We have meetings every other week. It's nice. We get to catch up, in terms of how I'm doing. If I'm struggling with stuff or if I need any help in anything, they're always there, and they're always encouraging." (junior undergraduate, 2018)

> "We have meetings every two weeks on every Wednesday at 12. Those help out sort of like a ...Well, it's not like a therapy session but for me, it kind of is a therapy session. I enjoy those. I look forward to those meetings." (junior undergraduate, 2018)

A primary function of Cal-Bridge mentoring is to show scholars the path to graduate education: to make them aware of the steps they should take, to support them in actually taking each step, and to keep them on track as they progress. **This function is especially valuable for students who are the first in their family to experience higher education:**

> "I didn't know much about what graduate school was like and what it took to get into a PhD program because my parents don't have the education background and I didn't have any idea. So one thing that was also really valuable for me besides the research experience is how the professors really care about you and they were always there for me. Because I always knew that I wanted to do science but I didn't know how to get there." (senior undergraduate, 2018)

> "They're always on you like, "Okay, have you done this, have you done that?" They're always checking on you. You're always meeting with the mentors every two weeks to check on your grades to see that you're on track. They know that us first-generation students, sometimes you don't really have that information, or you don't know how to get there, and so sometimes you need the extra push and extra hands to help you get there." (senior undergraduate, 2018)

> "The best part of the Cal-Bridge program is someone to tell me and show me everything to get into graduate school. They explain everything like I'm five, but when I was five no one ever explained the school hierarchy and paths to me." (junior undergraduate, 2020)

**Scholars expressed particular appreciation for mentors that not only provide academic support, but also show interest in their well-being as a whole person:**

> "He'll be like, okay, give me the rundown on all of your classes. And then he'll be like, okay, how are *you* doing? How are you managing things? Is there anything coming up that you're worried about? … They make a concerted effort to mention that I'm more than just school. ...I feel like it's a mixture of them wanting to make sure that I do well, but they're also making sure that *I'm* well. And I like that a lot." (senior undergraduate, 2020)

Cal-Bridge mentors are not necessarily members of underrepresented groups themselves; in 2012 only 5% of physics and astronomy faculty nationally were Black or Hispanic (Ivie et al. 2014) and numbers have not improved since then. That said, **when there is the opportunity for Cal-Bridge scholars to be paired with someone from their identity group, the relationship can be especially meaningful, even transformative:**

> "When I saw that I got another Black female to be my mentor, I literally broke down. I didn't know how much it would affect me… I looked her up and I was like, I can't handle this right now. I realized just how much that meant to me. So I've tried to put myself in that position for other people. I don't know who else needs that, because I didn't know that I needed it, so I want to make sure that I'm available to the rest of my community." (senior undergraduate, 2020)

When asked to rate the quality of five elements of the Cal-Bridge program across the 4 pillars (faculty mentorship, providing a path to graduate school, financial aid, workshops) and including a 5th element-the peer network, scholars rated faculty mentorship the highest, with over 50% of respondents rating this feature as excellent and over 90% rating mentorship as "Good" or better (Figure 3). All elements of the program were rated as "Good" or better by 75% or more of scholars. However, as much as scholars appreciated faculty mentorship and the other elements of the program, at the time of these interviews and surveys scholars were relatively dissatisfied with their opportunities for peer networking. Figure 3 indicates that the peer network was at that time the lowest-rated feature of the Cal-Bridge program. For example, one scholar from the first cohort stated,

> "To me, the only drawback is I never felt like we developed a good sense of community. I've connected to one good friend there. Everyone else, I don't know their number. Sometimes I forget their name." (junior undergraduate, 2018)

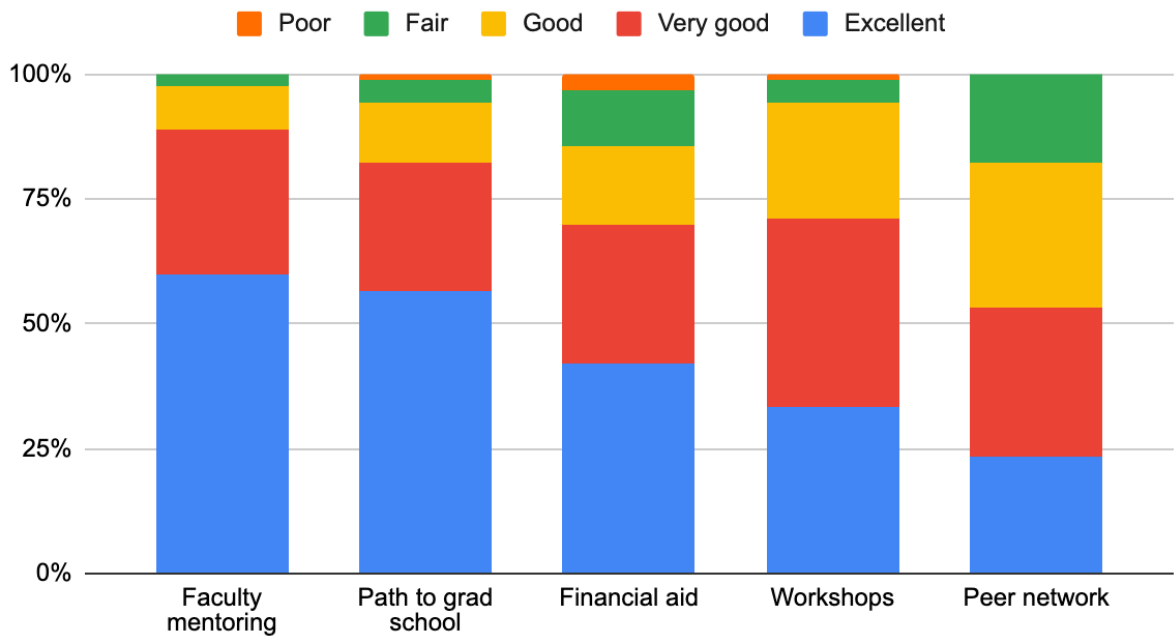

*Figure 4. Rating of quality of program elements based on survey responses in 2019-2020.*

# Application of findings

Our results highlight (1) the importance of faculty mentorship and (2) the desire for community amongst scholars. In response to these priorities, the program initiated two major changes, in the 2019-2020 academic year: (1) increased support for faculty and scholars to strengthen their mentoring relationships and (2) programmatic support for peer networking, the Peer Mentorship Program.

## Support to strengthen faculty mentorship

Scholars consistently reported that their relationships with faculty mentors are key to their success, particularly when mentors are caring and support the scholar's well-being as a whole person. These strongly positive experiences, alongside recent research emphasizing the importance of mentorship as a mutual relationship or working alliance (National Academies of Sciences, 2019), provided the impetus for the program to provide professional development to both faculty and scholars to strengthen and deepen mentoring relationships.

Mentorship practices for the Cal-Bridge program are documented in a Cal-Bridge Mentorship Handbook.[7] The Handbook was originally written by a small group of faculty mentors solely for other faculty mentors. The Handbook has since been transformed into a living document rewritten and edited by five scholar alumni (including author Rodriguez Wimberly), with input from program directors (including authors Rudolph and Hood); the intended audience is both faculty mentors and scholars. The new Handbook shares experiences and evidence-based best practices from both mentor and scholar perspectives, including:

- Discussion of struggles common to students with identities marginalized in STEM
- Suggestions to faculty mentors of ways to assist their scholar
- Support for scholars to tend to their mental health
- Support for scholars to seek assistance both from and beyond their two official faculty mentors
- Calls to action to increase equity in the academic system

A key addition to this new Handbook is the inclusion of specific guidance for a variety of cultural and identity-related topics. For example, a "Culture Shock" section shares the perspectives of international scholars and DACA/AB540 scholars,[8] written by alumni for whom these are their lived experiences.

The practices and advice documented in the Handbook are the basis for professional development provided to both scholars and mentors, aimed at supporting deeper and stronger mentoring relationships. At a workshop for faculty mentors, scholars, and alumni, participants used the Handbook to help them navigate example scenarios. This case-based approach to mentorship education has proven effective in other settings (Pfund et al., 2006, 2014). For example, in one (fictitious, but common) scenario, a scholar brings up feelings of isolation to their mentor and seeks guidance on how to overcome these feelings; discussion groups agreed that the faculty mentor in this scenario should support the scholar in taking time to prioritize their mental health and, in so doing, normalize and de-stigmatize mental health issues for scholars. In a more challenging scenario, a scholar shares with her faculty mentor that their previous mentor is rejecting their request for a letter of recommendation based explicitly on one of the scholar's underrepresented identities. Discussions of this example urged the faculty mentor to speak out against the other faculty member's inappropriate statements in an effort to improve academic culture, while also striving to minimize harm to the scholar. Throughout the workshop, faculty mentors, alumni scholars (graduate students) and current scholars (undergraduates) communicated equally and shared the responsibility of generating solutions. This highlights the collaborative nature of healthy mentorship — the main goal in rewriting Cal-Bridge's Mentorship Handbook inspired by the program's evaluation findings.

---

[7] The Handbook is publicly available at tinyurl.com/CBMentorshipHandbook and has been written in such a way as to be usable by any mentorship-centered program.
[8] DACA is the Deferred Action for Childhood Arrivals program which invites undocumented immigrants to work legally in the United States. The AB 540 statute exempts certain California students from paying nonresident tuition and allows them to apply for and receive state financial aid.

The current study is limited in its ability to inform effective mentoring practices. Mentor demographics were not collected, allowing for the possibility of patterns this study cannot detect (for example, whether women are especially effective mentors). Mentor professional development opportunities were offered to all mentors, but were not attended by all mentors. Future research may investigate the effects of professional development opportunities on mentorship effectiveness (as rated by mentees).

## Support to strengthen peer network

Peer mentorship groups have been shown to promote collaboration, provide both mentees and their near-peer mentors with both psychosocial and career support and increase retention (L. S. Tenenbaum et al., 2014). Peer or near-peer groups may also serve to enhance self-efficacy and diminish feelings of isolation: for example, the Fisk-Vanderbilt Master's-to-PhD Bridge Program has found that a tiered, peer mentorship approach involving seniors linked to first year students, helps students feel emotionally supported (Stassun et al., 2010). Unfortunately, the lowest rated element of the Cal-Bridge program as of 2020 was the Peer Network.

In response to this finding, Cal-Bridge developed a Peer Mentorship Program to strengthen the sense of community amongst current and former scholars and to broaden the scholars' mentor networks. Volunteer peer mentors (Cal-Bridge alumni scholars) are paired with mentees based on the mentee's individual preferences (e.g. academic seniority level, research area, race/ethnicity, gender). Pairs are self-guided in discussion content; meeting times and expectations are provided to mentors to set common communication, scheduling, follow through and code of conduct guidelines. Peer mentors also meet with scholars in workshops led by a Cal-Bridge alum or a graduate division counselor or professional development group from a UC campus. All workshops are highly interactive and optional with the curriculum chosen by scholars. For example, in one workshop activity, scholars work with the peer mentor to create a "mentor map"[9] identifying individuals or groups for whom they seek guidance, support or inspiration in different areas such as Accountability, Feedback, Self-Assessment, Health,

---

[9] The mentorship map is publicly available at tinyurl.com/CBMentorshipMap.

Intellect, Aspirations, Development and Opportunities. The "mentor map" was inspired by other mentor network worksheets (cf. Montgomery, 2017; the National Center for Faculty Diversity and Development Mentoring Map; the International Association of Women Mentor Map; and San Diego State University's Office of Faculty Advancement Mentor Map)[10] and was adapted to suit STEM undergraduates.

In its first 2 pilot years, the peer mentorship programming was rated positively as having added a valuable experience to the overall Cal-Bridge programming. Surveys were given to participants in the peer mentorship program in AY 19-20 and AY 20-21. In AY19-20 (AY20-21), there were 62 (71) active scholars, 18 (16) of whom expressed interest in participating as a mentee in the Peer Mentorship Program. 22 (20) scholars responded to the 2020 (2021) end-of-the-year evaluation survey. Figure 4 displays the ratings from these evaluation surveys.The 1-on-1 component was offered to scholars in Northern California only in AY 19-20 and all scholars the following year. In the first year, 71% (7 respondents) found the paired peer mentorship useful while the following year all mentees found meetings with their assigned peer mentor very beneficial. Similarly, the workshop component was offered to scholars in Southern California only in AY 19-20 and all Scholars the following year. In AY 19-20, 98% of survey respondents (15) felt the workshops had at least a minor positive value-added impact on their academic paths while in AY 20-21 this dropped slightly to 78% (20 respondents) finding positive impact. Over the course of the programming the sense of community, one of the main motivations for creating these mentorship opportunities, increased among participants. In the first year no more than 50% of survey respondents (22) felt the programming had a positive impact on the strength of their relationships with scholars and alumni. In AY 20-21, 95% of survey respondents (20) felt they had connections within the Cal-Bridge Scholar community for whom they could reach out to for help. As the program progresses more work will be put into defining the sense of community, increasing participation, and improving the workshops and 1-on-1 discussion structures to be of maximum benefit to scholars.

---

[10] See Acknowledgements Section for links.

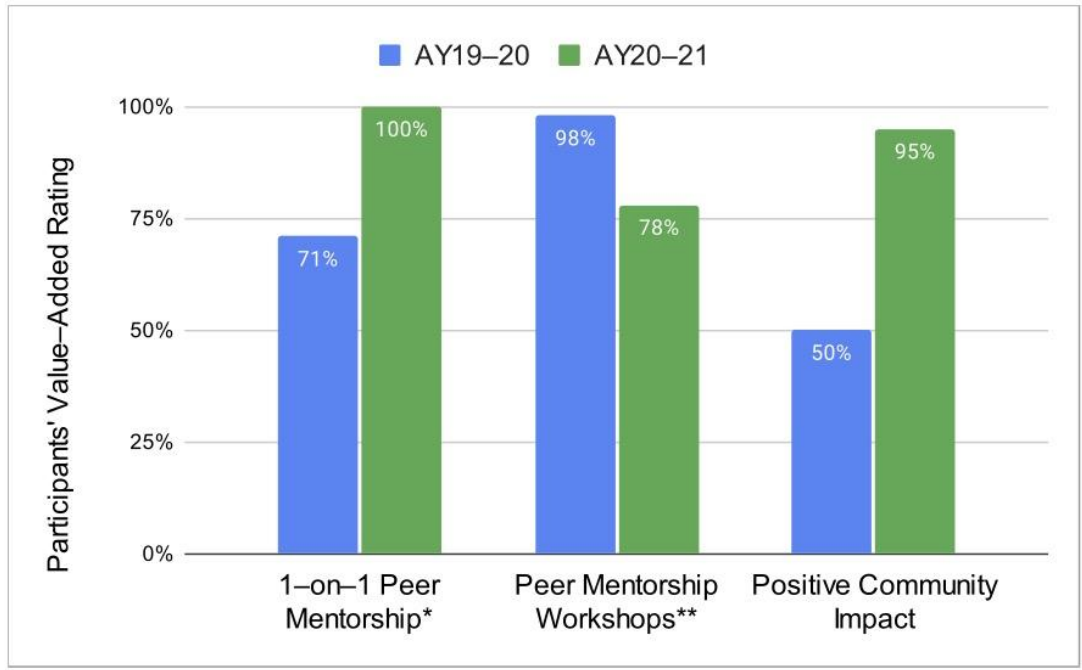

*Figure 5. Rating of the value added to the Cal-Bridge experience from Peer Mentorship Program elements based on survey responses in 2019-2021.[11,12]*

# Conclusion: A model of mentorship

Research has shown how critical mentorship and community-building are to student academic success and persistence, especially for students from historically underrepresented groups (see National Academies of Sciences, 2017, 2019; Pfund, 2018; and Rath et al., 2018 for a summary of the literature). In this work we detailed how the various elements of the Cal-Bridge program supported participating scholars, the majority of whom belong to historically underrepresented groups in their field of study. By far the most valuable element of the Cal-Bridge program as perceived by its scholars is faculty mentorship: scholars attribute their academic success to mentorship from their assigned faculty mentors.

Our data suggested additional avenues to strengthen faculty mentorship and areas for further program improvement, e.g., peer networking.  In response, Cal-Bridge-designed faculty professional development to support deepening mentor-scholar relationships, and created peer networking opportunities to connect current and alumni scholars. Impacts of the Peer Mentorship Program will be the subject of an upcoming paper (Rodriguez Wimberly, *in prep*). Additionally, future work will explore the effects of mentor professional development opportunities on mentorship effectiveness, as rated by mentees. As Cal-Bridge expands to support graduate students and postdoctoral scholars, vertical mentoring relationships will be formalized to broaden the *relationship-centered model of mentorship* to include all academic levels in this crucial support structure.

---

[11] *The 1-on-1 Peer Mentorship element was only offered to Cal-Bridge scholars in Northern California in AY19-20.*
[12] *The Peer Mentorship Workshops element was only offered to Cal-Bridge scholars in Southern California in AY19-20.*

Results of this program highlight a *relationship-centered model of mentorship* as being critical in the success of young academics, especially those with identities that are underrepresented in their field. The features of this model are:
- Multiple mentors provide a network of support
- Mentorship education includes interactive, group elements and resources
- Continuous improvement is supported by quantitative and qualitative assessments

Together these features created the *multifaceted highly-engaged mentoring relationships* which are most successful in broadening the participation of STEM students with historically underrepresented backgrounds.

# Acknowledgements

The Cal-Bridge Mentorship Map, discussed in the Application of Findings: Support to strengthen peer network section, was created through inspiration by the following mentor maps: Beronda L. Montgomery's "Mapping a Mentoring Roadmap and Developing a Supportive Network for Strategic Career Advancement" (https://doi.org/10.1177/2158244017710288), the National Center for Faculty Diversity and Development Mentoring Map (https://www.facultydiversity.org/ncfddmentormap), the International Association of Women Mentor Map (https://info.iawomen.com/hubfs/Mentor%20Map.pdf) and San Diego State University's Office of Faculty Advancement Mentor Map (https://newscenter.sdsu.edu/facultyadvancement/files/06543-Mentor_Map.pdf).

This material is based upon work generously supported by the National Science Foundation under Grant No. DUE-1741863. MKRW acknowledges support from the National Science Foundation MPS-Ascend Postdoctoral Research Fellowship under Grant No. AST-2138144. CP acknowledges support from the NSF INCLUDES Alliance: Inclusive Graduate Education Network Grant No 1834540.

# Appendix A: Interview protocol

To help create an environment with the best opportunity for shared understanding, the interviewer took the following steps:

- The interviewer explicitly invited interviewees to decline any question that they preferred not to answer.
- The interviewer first asked interviewees questions about their experience in their undergraduate program, without reference to race or ethnicity: how their classes and research were going, what their relationships with faculty were like, and so on. The intention was to allow students to speak about their experiences from whatever perspective felt primary to them, without priming them.
- Later in the interview, interviewers explained the reasons for asking questions about interviewees' racial or ethnic experiences and asked for their consent to share experiences from their perspective as a member of an underrepresented group. The intention was to invite and endorse students' speaking from their perspective as a member of an underrepresented group.
- The interviewer invited interviewees to describe their racial or ethnic identity in their own terms and used interviewees' preferred descriptors, rather than imposing categories defined by the Cal-Bridge program or other entities.

The detailed interview protocol is given in Table A1.

Table A1. Interview protocol

1. How is school going for you? (classes, research)
2. How is your life outside of school? (work outside of school, social life, life events)
3. You may have interactions with people who have mentored or advised you. Could you tell me who those people are for you? Do you talk about academic matters, or personal matters? Is this relationship what you expected? What kind of benefit do you get from this mentoring? Do you have a trusting relationship with this mentor? What do you think led to it being a trusting relationship? (These questions are repeated for each mentor the interviewee prioritizes. Prompt specifically for their Cal-Bridge mentors if they do not bring them up on their own.)
4. Do you consider any other Cal-Bridge scholars to be "peer mentors" for you? Have you participated in the peer-mentoring program? Are you connected to other Cal-Bridge scholars?
5. One goal of the Cal-Bridge program, as you may know, is to increase the number of physics PhDs from underrepresented groups. The program would like to know more about the experiences of students from underrepresented groups in physics and astronomy, so that the program can better support people in the future.
   a. Do you identify as a member of an underrepresented group in physics? How would you describe yourself? [Answer = descriptor, below] If yes: Would you be willing to share your experiences from your perspective as [descriptor] to help us answer this question?

b. (If yes: ) Do you think that being [descriptor] matters for your experience in physics / astronomy? If so, how?

c. (If yes: ) Do you think that being [descriptor] matters for your mentoring relationships? If so, how?

d. Are there any other [descriptor] students in your department (that you know of)?

e. Are you connected with a group of [descriptor] in physics? How/where?